\documentclass[12pt, onecolumn]{article}

\newcommand{\newc}{\newcommand}
\newc{\gsim}{\lower.7ex\hbox{$\;\stackrel{\textstyle>}{\sim}\;$}}
\newc{\lsim}{\lower.7ex\hbox{$\;\stackrel{\textstyle<}{\sim}\;$}}
\newc{\gev}{\,{\rm GeV}}
\newc{\mev}{\,{\rm MeV}}
\newc{\ev}{\,{\rm eV}}
\newc{\kev}{\,{\rm keV}}
\newc{\tev}{\,{\rm TeV}}

\newc{\mz}{M_Z}
\newc{\mpl}{M_*}
\newc{\mw}{m_{\rm weak}}
\newc{\nr}[1]{N^c_R{}_{#1}}

\def\beq{\begin{equation}}
\def\eeq{\end{equation}}
\def\bea{\begin{eqnarray}}
\def\eea{\end{eqnarray}}
\def\bitem{\begin{itemize}}
\def\eitem{\end{itemize}}
\newc{\ie}{{\it i.e.}}          \newc{\etal}{{\it et al.}}
\newc{\eg}{{\it e.g.}}          \newc{\etc}{{\it etc.}}
\newc{\cf}{{\it c.f.}}
\def\bar#1{\overline{#1}}
\def\vev#1{\left\langle #1 \right\rangle}

\def\abs#1{\left| #1\right|}

\def\inv{^{\raise.15ex\hbox{${\scriptscriptstyle -}$}\kern-.05em 1}}
\def\lbar{{\lower.35ex\hbox{$\mathchar'26$}\mkern-10mu\lambda}} %lambda bar

\let\<=\langle
\let\>=\rangle

\let\+=\uparrow
\let\al=\alpha
\let\be=\beta
\let\ga=\gamma

\let\ep=\epsilon
\let\ze=\zeta

\let\la=\lambda

\begin{document}
\thispagestyle{empty}
\vspace*{.5cm}
\noindent
\hspace*{\fill}{\large OUTP-04/16}\\
\vspace*{2.0cm}

\begin{center}
{\Large\bf Naturally Degenerate Right Handed Neutrinos}
\\[2.5cm]
{\large Stephen M. West
}\\[.5cm]
{\it Rudolf Peierls Centre for Theoretical Physics,\\
University of Oxford, 1 Keble Road, Oxford OX1 3NP, UK}
\\[.2cm]
(August, 2004)
\\[1.1cm]

{\bf Abstract}\end{center}
\noindent
In the context of supersymmetric theories, a weakly broken gauged SO(3)
flavour symmetry is used to produce two highly degenerate right handed
(RH) neutrinos. It is then shown that this SO(3) flavour symmetry is
compatible with all fermion masses and mixings if it is supplemented
with a further SU(3) flavour symmetry. A specific Susy breaking model
is used to generate the light neutrino masses as well as a natural
model of $\tev$ scale resonant leptogenesis.

\newpage

\setcounter{page}{1}

\section{Introduction}

$\tev$ scale leptogenesis is an important alternative to the
leptogenesis model associated with the seesaw mechanism
\cite{FY}. The standard see-saw mechanism \cite{seesaw} prescribes
heavy RH neutrinos and it is the decay of these states that
can lead to an asymmetry in lepton number. At this high scale
the Hubble constant, H, is generally larger than the decay widths of the RH
neutrino states and consequently they decay out of thermal
equilibrium. This departure from thermal equilibrium ensures that any
asymmetry produced is not immediately washed out by inverse decays or any
scatterings that involve the RH neutrino.  However, due to the high
mass scale of the RH neutrinos the see-saw mechanism and its'
associated leptogenesis mechanism are difficult to directly test. This
is in contrast to $\tev$ scale theories of neutrino mass generation
and leptogenesis \cite{Flanz,CRV,Pil1,TH,Pil2,HLNPS,hmw,bhs,tev,pil3}.
One of the more attractive features of low scale theories is the
possibility of being able to directly test components of the model.

A TeV scale theory will have a small Hubble constant. We require
that the various scatterings which can suppress an asymmetry be
under control. At these low scales gauge scatterings are very fast,
consequently a singlet of all low energy gauge symmetries is
preferred for the decaying particle. Considering standard thermal
leptogenesis, one can think about various possibilities with
decaying singlet particles at low scales: a large degeneracy of
masses between the decaying particles
\cite{Flanz,CRV,Pil1,TH,Pil2,HLNPS,hmw,pil3}; a hierarchy between
the couplings of real and virtual particles in the one loop
leptogenesis diagrams \cite{TH,bhs}; or three body decays of the
heavy particles with suppressed two body decays \cite{TH} (for
related work in leptogenesis see, \cite{others,krystoff}).

In this letter we will concentrate on the possibility of decaying $\tev$
scale RH neutrinos with a large degeneracy in their masses. This %%@
framework suffers from various significant difficulties:

1) Seesaw type neutrino masses require tiny couplings and
consequently will usually induce a tiny CP asymmetry.

2) We need the decay width of the particle which generates the
asymmetry to be less than H, so that the particle
decay will be out of thermal equilibrium and any asymmetry produced
is not immediately washed out. This again requires tiny Yukawa
couplings of order $ 10^{-6}-10^{-7}$. Such small couplings need
justification.

3) In a generic seesaw model there is no explanation why the RH
neutrinos would have such a small mass ($M_N\sim$~$\tev$).

4) In order to compensate the large suppression
of the asymmetry induced by these tiny couplings, an extremely tiny mass %%@
splitting is required between two RH neutrino masses giving a
resonant behaviour in the RH neutrino propagator. The degree of
degeneracy required has to be of order
$(M_{N_1}-M_{N_2})/(M_{N_1}+M_{N_2}) < 10^{-10}$ \cite{Pil2}. This
level of degeneracy needs to be physically motivated.

5) Finally, as a result of the constraints 1) and 2) the tiny Yukawa
couplings imply that the RH neutrino production cross sections are
very small. Which means that even at low scales the theory may not be testable.

In this letter we will argue, extending the arguments of
\cite{hmw,gmnu,mw}, that these potential difficulties can be
overcome. In the context of broken susy Ref. \cite{hmw} considered
two or more quasi-degenerate RH neutrinos. In this case the
asymmetry can be significantly enhanced through a resonant behaviour
of the propagator of the virtual particle in the leptogenesis
self-energy diagram \cite{hmw}. This model possesses a natural
explanation for both tiny Yukawa couplings and $\tev$ scale RH
neutrinos (see Ref \cite{mw} for more details). Now one would like
to form a natural explanation for the high degree of degeneracy in
the RH neutrino spectrum.

In the following section we propose an SO(3) flavour symmetry which
can be used to produce two exactly degenerate RH neutrinos
\footnote{An SO(3) symmetry has been previously used in connection
  with quasi-degenerate light neutrinos, see Ref.\cite{so3}.
}. In  Section 3 a toy model is outlined where the SO(3) flavour
symmetry is embedded into the susy breaking model described in 
Refs.\cite{hmw} and \cite{mw}. Utilising a further SU(3) flavour
symmetry it is shown that
all fermionic standard model sectors including neutrino masses and
mixings are compatible with this SO(3) flavour symmetry\footnote{In this paper we want
 to argue that there exists a model with naturally degenerate RH
 neutrinos justified by a symmetry,
  which is compatible with the Standard Model. It is not claimed that
  this is the most minimal solution.}. Following
this we go on to describe a natural and successful model of $\tev$ scale
resonant leptogenesis. Our conclusions are contained in Section 4,
while two appendices contain technical details of the models presented.

\section{The SO(3) flavour symmetry}

We assume the minimal supersymmetric standard model (MSSM) with the
addition of standard model singlet RH neutrino chiral
supermultiplets, $N_i$. Under a gauged SO(3) flavour symmetry
$N_{i}$ transforms as a triplet, where  $i$=1,2,3 (and all other
Roman indices) are SO(3) labels. All other MSSM chiral
supermultiplets transform as singlets under this SO(3) flavour
symmetry.

We need to spontaneously break the SO(3) flavour symmetry \footnote{Using a
gauged SO(3) symmetry means that any potentially dangerous massive
vectors are avoided.}. This is performed by two flavon
fields, $\ze$ and $\xi$, developing vacuum expectation values
(VEVs). Each field is a triplet under the SO(3) flavour symmetry but a
singlet under the standard model gauge group.

\subsection{Degenerate Right Handed Neutrinos}

RH neutrino masses can be generated via the superpotential or the
Kahler potential depending on how
exactly the scale of their masses is realised. This letter concentrates on
the generation of $\tev$ scale RH neutrinos via non-renormalisable
operators arising from the Kahler potential. However, as a simple example of
how the SO(3) flavour symmetry can generate degenerate RH neutrinos it
is appropriate to study the mechanism in the context of an effective
superpotential. Using the flavon field discussed above we can write down,
\beq
M_N  \int d^2\theta \left( h_1 N_{i} N_{i} + \frac{1}{M_{f}^2}
h_{2} N_{i} \ze_{i} N_{j} \xi_{j} + \frac{1}{M_{f}^4}
  h_3 \ep_{ijk}N_{i} \ze_{j} \xi_{k} \ep_{lmn}N_{l}
\ze_{m} \xi_{n}\right) \label{suprhn} \eeq
where $\ep_{ijk}$ is the usual antisymmetric tensor, M$_N$ is the
RH neutrino scale, M$_f$ is the cut off scale, which we assume is the
mass scale of some heavy
fields that have been integrated out, all $h$s 
are undetermined 
{\it O}\hspace{0.5mm}(1) parameters and we assume the R-parities of
$\ze$ and $\xi$ are equal in magnitude but opposite in sign.

We assume the two flavon fields develop VEV structures given as,
\beq
 \vev \ze  =
 \left(\begin{array}{ccc}
   A \\
   iA \\
   0
\end{array}\right) \hspace{2 mm} , \hspace{10 mm} \vev \xi =
 \left(\begin{array}{ccc}
   D \\
   -iD \\
   0
\end{array}\right)
\eeq where $A$ and $D$ are related and can be complex. The alignment
of these two VEVs is crucial for the generation of degenerate RH
neutrinos and is presented in the next section. It is assumed that
the VEVs of $\ze$ and $\xi$ are comparable to the high scale so that
$a$ $\equiv$ $A/M_{f}$ and $d$ $\equiv$ $D/M_{f}$ are not much
less than one.

Allowing the two fields to acquire their VEVs the RH neutrino
mass matrix has the form,

\beq M^{sp}_N \sim \left(\begin{array}{ccc}
 h_1 + h_2 ad  & 0  & 0 \\
   0 &   h_1 +h_2 ad  & 0 \\
 0 & 0  &  h_1 +h_3 4 a^2 d^2
\end{array}\right)
\eeq where a minus sign has been absorbed into the definition of
$h_3$. There are further terms that can be written down in
addition to those in equation (\ref{suprhn}) but none of these give
either non-diagonal or differing (1,1), (2,2) entries in the mass
matrix. Consequently we produce two exactly degenerate RH neutrinos
\footnote{In this example we have no constraints on the sizes of $a$ and $d$,
  but for $ad > 1/4$ we give $N_3$ a larger mass than $N_1$ and
  $N_2$. Consequently resonant leptogenesis could proceed via the decay
  of $N_1$ and $N_2$.}.

\subsection{Vacuum Alignment}
The crucial part of this model is the vacuum alignment which
determines the structure of VEVs for the fields $\ze$ and $\xi$.
This section will discuss how exactly this alignment can arise. The
first stage of the symmetry breaking is triggered by the $\ze$ field
acquiring a VEV radiatively. We assume that the soft mass of the
$\ze$ field gets driven negative at some scale through radiative
corrections. This could be achieved if we assume the field $\ze$ has
Yukawa couplings to a massive field. Such radiative effects can
trigger a VEV for $\ze$ \cite{kr}. We have the freedom to rotate the
VEV of $\ze$ to read $\vev \ze^T = \left(A, B, 0 \right)$ without
loss of generality. At this point there is nothing to say whether
$\xi$ gets a VEV or not so we assign an arbitrary structure to $\xi$
of the form $\vev \xi^T = \left(D,E,F\right)$, where $D$,$E$ and $F$
can still be zero. The superpotential terms \beq S\sim  P\ze_i \ze_i
+ T\xi_j \xi_j \label{PT} \eeq can be written down assuming
consistent R-charge assignments (a specific example is given in
later sections and in Appendix B). Along the F-flat direction
$|F_P|^2=0$, we have $\langle \ze^2 \rangle=0$, which forces
$A=-iB$, leading to $\vev \ze^T = \left(A, Ai, 0 \right) $.
Moreover, along the F-flat direction $|F_T|^2=0$ we have the
condition, \beq \vev {\xi^2} = D^2+E^2+F^2=0 \label{fxi} \eeq

In order to have radiative corrections generating large VEVs they must
evolve along D-flat directions. The conditions for D-flatness arising
from the generators \beq T_1 =
\frac{1}{2}\left(\begin{array}{ccc} 0 & 0  & 0 \\
   0 & 0 & -i \\
 0 & i & 0
\end{array}\right), \hspace{5mm}
T_2 = \frac{1}{2}\left(\begin{array}{ccc}
0 & 0  & i \\
   0 & 0 & 0 \\
 -i & 0  & 0
\end{array}\right), \hspace{5mm}
T_3 = \frac{1}{2}\left(\begin{array}{ccc}
0 & -i  & 0 \\
   i & 0  & 0 \\
 0 & 0  & 0
\end{array}\right)
\eeq
are of the form
\bea
\abs{D_1}^2 & \propto & \abs{EF^*-E^*F}^2=0 \label{d1}\\
\abs{D_2}^2 & \propto & \abs{FD^*-F^*D}^2=0 \label{d2} \\
\abs{D_3}^2 & \propto & \abs{2\abs{A}^2+i(DE^*-D^*E)}^2=0 \label{d3}
\eea
A solution to conditions (\ref{d1}) and (\ref{d2}) is $F=0$. Applying
this condition to (\ref{fxi}) and rewriting the potentially complex
parameters $D$ and $E$ as $D=D_R+iD_I$ and $E=E_R+iE_I$ we have,
\bea
D_R^2-D_I^2+E_R^2-E_I^2=0 \label{dm1} \\
D_RD_I+E_RE_I=0 \label{dm2}
\eea
 and (\ref{d3}) gives
\beq
\abs A^2=D_IE_R-D_RE_I \label{dm3}.
\eeq
Solving conditions (\ref{dm1}), (\ref{dm2}) and (\ref{dm3}) we are led
to the relations
\beq
D_R=-E_I , \hspace{5mm} D_I=E_R \hspace{5mm} \Rightarrow \hspace{5mm} %%@
E=-iD.
\eeq
Which means,
\beq
D_I=\pm \sqrt{\abs A^2 - D_R^2}
\eeq
where $-\abs A \leq D_R \leq \abs A$.
Finally the full expression  for $\vev \xi$ is
\beq
\vev \xi =
 \left(\begin{array}{ccc}
   D_R\pm i\sqrt{\abs A^2 - D_R^2} \\
   \pm \sqrt{\abs A^2 - D_R^2}-iD_R \\
   0
\end{array}\right)=
  \left(\begin{array}{ccc}
   D \\
   -iD \\
   0
\end{array}\right).
\eeq
Substituting these relations back into (\ref{d1}) and (\ref{d2}) we find $F=0$
is a consistent solution \footnote{In this analysis, possible soft mass terms
  for the flavon fields have been neglected. If we include such terms,
  we will generate corrections to the vacuum alignment above which are
  parametrically the scale of the soft masses.
  We expect these corrections to be of order $\sim M_{susy}$. When we
  include these corrections into the VEVs of $\ze$ and $\xi$ we generate
  non-diagonal and differing (1,1) and (2,2) terms in the mass matrix
  of the RH neutrinos of order $\sim M_{susy}^2/M_{f}$ at most.}.

\section{A Toy Model}

The aim of this section is to show that the SO(3) flavour
symmetry can be used in a model that successfully describes all
fermionic sectors including the generation of neutrino masses. We do
this using, along side the SO(3) flavour symmetry, an adaptation of
the model described in Ref.\cite{kr}. In this paper all the
MSSM fields including the RH neutrino field are triplets under an SU(3) %%@
flavour symmetry. However in our adaptation the RH neutrino fields are now %%@
singlets under the SU(3) flavour symmetry and a triplet under the new SO(3) %%@
flavour symmetry. The other MSSM fields are singlets under the SO(3) flavour
symmetry. Summarising, the flavour symmetry assignments we have for the %%@
SO(3) symmetry,

\beq (Q,L,U^c,D^c,E^c)\sim 1, \hspace{5mm} N_i\sim 3 \eeq and for
the SU(3) symmetry \beq (Q_\al,L_\al) \sim 3, \hspace{5mm}
(U^c_\al,D^c_\al,E^c_\al)\sim3, \hspace{5mm} N \sim 1 \eeq where
$\al=1,2,3$ (and all other Greek indices) are SU(3) labels. Moreover,
all Higgs fields responsible for SU(3) symmetry breaking as well as
any other fields used to achieve the desired vacuum alignment are
singlets under the new SO(3) flavour symmetry. A summary of all the
assignments is given in Appendix A. We use the mechanisms presented in
Ref.\cite{kr} for all sectors apart from the neutrino sector which
we present here.

\subsection{Neutrino masses from Susy breaking}

We need to generate neutrino masses and we do this in a similar way
to Ref.\cite{mw}. As emphasized by the authors of Ref.\cite{gmnu}, we
can apply the Giudice-Masiero mechanism \cite{gm} to the neutrino sector,
i.e SM-singlet operators, such as the RH neutrino mass $M_R N N$, or
the neutrino Yukawa coupling $\lambda L N H_u$, might only appear to
be renormalizable superpotential terms, but in fact may arise from
$1/M$-suppressed terms involving the fundamental supersymmetry
breaking scale $m_I \sim \sqrt{M_{3/2}M_{pl}}$, where
$M_{pl}$ and $M_{3/2}$ are the reduced Planck mass and gravitino mass
respectively.

Specifically, consider the usual MSSM Lagrangian to be supplemented
by Standard-Model-singlet chiral superfields which arise from the
hidden sector.  In general, the fields which communicates
supersymmetry breaking to the neutrinos can either be flavour
singlets or flavour non-singlets. Here we assume that all such
fields are singlets under all flavour symmetries.

Ignoring flavour and consequently suppressing all indices for the
moment, the scales of the various terms we wish to study are set by
the hidden sector fields acquiring VEVs. In the superpotential we
have \beq \mathcal{L}_N^W = \int d^2\theta \left( g\frac{T}{M} L N
H_u  \right), \label{newW} \eeq while the set of terms involving the
RH neutrino fields in the Kahler potential are \beq \mathcal{L}_N^K =
\int d^4 \theta \left( h \frac{T^\dagger}{M} N N + {\tilde h}
\frac{T^\dagger T}{M^2} N^\dagger N + h_B \frac{T^\dagger
TT^{\dagger} }{M^3} N N + \ldots \right). \label{newK} \eeq Here T
is a susy breaking hidden sector field and the ellipses in
(\ref{newK}) stand for terms higher order in the $1/M$-expansion. It
is simple to check the additional terms will lead to trivial or
sub-dominant contributions not relevant for our discussion. All
dimensionless couplings $g,h$, etc, are taken to be {\it O}\hspace{0.5mm}(1).

Let us now suppose that after supersymmetry is broken in the hidden
sector at the scale $m_I$, the field $T$ acquires the following $F$- and
$A$-component VEVs, \bea
&&\langle T\rangle_F = F_t = f_t m_I^2 \nonumber\\
&&\langle T\rangle_A = A_t = a_t m_I. \label{yvev} \eea Here $f_{t}$
and $a_{t}$ are {\it O}\hspace{0.5mm}(1).
Substituting these VEVs into Eq. (\ref{newW}) and (\ref{newK}) shows
that after susy breaking we produce; (1) the scale for neutrino Yukawa as
$\sim 10^{-7}-10^{-8}$, (2) RH neutrino mass scale at a $\tev$, (3) a trilinear
scaler A-term at a $\tev$, (4) RH sneutrino lepton-number violating
$B$-term with magnitude $B^2 \sim ({\rm few}\times
100\mev)^2$. We produce two sources of neutrino masses, a tree level
(see-saw) contribution as well as a dominant 1-loop
contribution, (\cite{gmnu}, \cite{mw}). In the next section we
outline how one could combine the susy breaking model described above with the
flavour symmetries, SO(3) and  SU(3) to give neutrino masses and
mixings compatible with current experimental bounds.

\subsubsection{RH Neutrino Mass Matrix}

In the susy breaking model described above the RH neutrino mass terms
arise from non-renormalisable Kahler potential operators. In order to
produce degenerate RH neutrinos this way consider,

\bea K \sim \frac{T^{\dagger}}{M_{pl}}\left(
h_4 N_i N_i+\frac{1}{M_{f}^2} h_5 N_{i} \ze_{i} N_{j}
\ze_{j}^*+\frac{1}{M_{f}^2} h_6 N_{i} \xi_{i} N_{j} \xi_{j}^*\right)\\
+\frac{T^{\dagger}}{M_{pl}}\left(h_7\frac{1}{M_{f}^4}
\ep_{ijk}N_{i} \ze_{j} \xi_{k} \ep_{lmn}N_{l}
\ze_{m}^* \xi_{n}^* + \dots\right)\label{n3n3}
\eea
where the ellipses represent further terms that do not contribute to 
non-diagonal terms or give differing
(1,1), (2,2) entries. We assume the R-charge assignments in Table 1 of
Appendix A. Allowing the flavon fields to gain their appropriate VEVs, the RH
neutrino mass matrix takes the following form,
\beq M_N \sim \left(\begin{array}{ccc}
 h_4+ h_5\abs a^2+h_6 \abs d^2  & 0  & 0 \\
   0 & h_4+ h_5\abs a^2+h_6 \abs d^2   & 0 \\
 0 & 0  &  h_4 +h_7'\abs a^2\abs d^2
\end{array}\right),
\label{RHnemass} \eeq generating two exactly degenerate RH neutrinos. $h_7'$ represents the fact that there are numerous terms of 
the same order as the term in (\ref{n3n3}) contributing to the
mass\footnote{ In order to be consistent with neutrino masses 
  and mixings, we take parameter values $a=d=0.4$. Even with these
  values the mass of $N_3$ is larger than that of $N_1$ and $N_2$ due
  to these additional terms.} of $N_3$.

\subsubsection{Trilinear Scaler $A$-term}
A very important term which contributes to the 1-loop neutrino masses
is the trilinear scalar A-term. The structure
of this term comes from the following leading order superpotential
operators, \bea S_A\sim \frac{T}{M_{pl}}\left(g_{1}\frac{1}{M_f^4}\ep_{ijk}N_i\ze_j\xi_k\frac{1}{M_3^7}L_\al\phi_{3}^\al(\bar\phi_{3}\phi_{3})^3(\ze \xi)\right)
\\ + \frac{T}{M_{pl}}\left(g_{2}\frac{1}{M_f^2}\ep_{ijk}N_i\ze_j\xi_k\frac{1}{MM_3^8}L_\al\phi_{23}^\al(\bar\phi_{3}\phi_{3})^4\right)
\\ + \frac{T}{M_{pl}}\left(g_{3}\frac{1}{M_f^4}\ep_{ijk}N_i\ze_j\xi_k \frac{1}{M
M_3}\ep^{\al\be\ga}L_\al\bar\phi_{23,\be}\bar\phi_{3,\ga}(\ze \xi)+
\dots \right)\eea
Giving the structure, \beq A_\nu \sim \left(\begin{array}{ccc}
 0 & 0 & 0 \\
   0 &  0 & 0 \\
 g_{3}4a^2d^2\ep i &  g_{2}2ad\ep i & g_{1}4a^2d^2 i
\end{array}\right)
\label{Aterm} \eeq where we have written
$\ep=b/M$ and $\ep$, $a$ and $d$ are expansion parameters. Here we
assume that the $\ep$ parameter can be different to the expansion
parameter for the up quark sector. The neutrino sector is generated via
non-renormalisable susy breaking operators, with the RH neutrino
transforming as a singlet under the SU(3) flavour symmetry in contrast
to Ref. \cite{kr} where the expansion parameters are identical for
the two sectors.

\subsubsection{Neutrino Yukawa term}
In order to generate neutrino masses and mixings it is necessary to
add two hidden sector superfields, $Z_1$ and $Z_2$, with properties
and charge assignments as listed in Table \ref{tab1} of Appendix
A. Specifically we assume the $Z$ fields gain A-component VEVs,
$\langle Z\rangle_A = A_z = a_z m_I$, with zero (or tiny)
F-component VEVs.

The Yukawa flavour structure has a contribution from the new fields,
$Z_1$ and $Z_2$ in addition to a contribution from the field $T$.
The contribution from the field $T$ has exactly the same structure
as the trilinear scaler A-term except for the Yukawa the A-component
VEV of T is used. Leading order contributions from fields $Z_1$ and
$Z_2$ are, \bea S_{Yuk}\sim
\left(\frac{Z_1}{M_{pl}}g_{4}\frac{1}{M_f}N_i\ze_i+\frac{Z_2}{M_{pl}}g_{7}\frac{1}{M_f}N_i\xi_i\right)\frac{1}{MM_3^{10}}L_\al\phi_{23}^\al(\bar
\phi_3\phi_3)^5 \\ +
\left(\frac{Z_1}{M_{pl}}g_{5}\frac{1}{M_{f}}N_i\ze_i+\frac{Z_2}{M_{pl}}g_{8}\frac{1}{M_{f}}N_i\xi_i\right)\frac{1}{M_3^9M_{f}^2}L_\al\phi_{3}^\al
(\bar \phi_3\phi_3)^4(\ze\xi) \\
+\left(\frac{Z_1}{M_{pl}}g_{6}\frac{1}{M_{f}}N_i\ze_i+\frac{Z_2}{M_{pl}}g_{9}\frac{1}{M_{f}}N_i\xi_i\right)
\frac{1}{M
M_3^3M_{f}^2}\ep^{\al\be\ga}L_\al\bar\phi_{23,\be}\bar\phi_{3,\ga}(\bar
\phi_3\phi_3)(\ze\xi).\label{syuk}\eea
Giving the leading order Yukawa structure,\beq \left(\begin{array}{ccc}
 (a_{Z1}g_{6}a+a_{Z2}g_{9}d)2ad\ep &  (a_{Z1}g_{4}a+a_{Z2}g_{7}d)\ep  & (a_{Z1}g_{6}a+a_{Z2}g_{9}d)2ad \\
 (a_{Z1}g_{6}a-a_{Z2}g_{9}d)2iad\ep &   (a_{Z1}g_{4}a-a_{Z2}g_{7}d)\ep
 i & (a_{Z1}g_{6}a-a_{Z2}g_{9}d)2iad\\
 g_{3}a_T 4a^2d^2\ep  i &  g_{2}a_T2ad\ep i &  (g_{2}\ep+g_{1}2ad)a_T2ad  i
\end{array}\right).
\label{Yuk} \eeq

\subsubsection{Other Terms of Note}

The lepton-number violating $B$-term is crucial to the formation of
the 1-loop contribution to the light neutrino masses. The structure
of the $B$-term assuming $a$ and $d$ are real for simplicity, is
\beq \left(\begin{array}{ccc}
(h_4+ h_5  a^2+h_6 d^2)a_t  & 0  & h_{16}ia^2d(a_{z1}+h'_{16}a_{z2}) \\
0 &  (h_4+ h_5 a^2+h_6 d^2)a_t & h_{16}a^2d(a_{z1}-h'_{16}a_{z2}) \\
h_{16}ia^2d(a_{z1}+h'_{16}a_{z2}) &
h_{16}a^2d(a_{z1}-h'_{16}a_{z2}) &  (h_4 + h_8)a_t
\end{array}\right)
\label{bterm} \eeq which we generate from operators of the form of
the 3rd term in equation (\ref{newK}) and similar operators with one
of the $T^\dagger$s being replaced by a $Z^\dagger$.

We can also generate small corrections to the RH neutrino mass matrix
using the same form of operator. This is achieved when $T^\dagger$
gets an F-component VEV and two other hidden sector fields get
A-component VEVs. (The other two hidden fields could be $T^\dagger
T$,$Z^\dagger Z$,$T^\dagger Z$ or $Z^\dagger T$.) The resulting
structure of this splitting term, $\Delta M_{N}$, in the limit where $a \sim d$, \beq
\left(\begin{array}{ccc}
(h_4+ h_5  a^2+h_6  d^2)a_t  & 0  & ia^3a_3^2(a_{z1}h_{18}+a_{z2}h_{18}) \\
0 &  (h_4+ h_5  a^2+h_6 d^2)a_t &a^3a_3^2(a_{z1}h_{18}-a_{z2}h_{18}) \\
ia^3a_3^2(a_{z1}h_{18}+a_{z2}h_{18}) & a^3a_3^2(a_{z1}h_{18}-a_{z2}h_{18})   &  (h_4 + h_8)a_t
\end{array}\right)
\label{split} \eeq with a scale of $\sim 10^{-13}$ GeV and where
numerical factors have been ignored.
These splittings actually play no significant role in splitting of the
RH neutrinos as they enter into the matrix as mixings between the 1st
and 3rd and 2nd and 3rd generations.

\subsection{Neutrino Masses and Mixings}
As is described in Ref.\cite{mw} neutrino masses can be generated
from two different sources. The dominant piece is that produced
by a 1-loop contribution. The flavour structure of this contribution
in the limit that there is no mixing in the sneutrino sector is,
\beq
m_\nu^{\rm loop} \sim
A^TB^*A \eeq
Substituting in the forms for A and B from Eqs. (\ref{Aterm}) and (\ref{bterm})
respectively we get the structure,
\beq
m_\nu^{\rm loop} \sim a^2d^2 \left(\begin{array}{ccc} ad\ep^2 &
    ad\ep^2 & ad\ep^2+a^2d^2\ep
    \\ ad\ep^2 & \ep^2  & \ep^2+ad \\
ad\ep^2+a^2d^2\ep & \ep^2+ad & \ep^2+a^2d^2+ad\ep
\end{array}\right)
\eeq where numerical factors and various h and g coefficients have
been suppressed for simplicity. The form of this neutrino mass can
be identified with the structure for a normal hierarchy of neutrino
masses. On its own it can successfully generate the atmospheric
neutrino mass data. However in its current form it is rank 1. We now
need the second source of neutrino masses which comes from the tree
level ``see-saw'' contribution. This has the form,
 \beq
(m_\nu^{\rm tree})_{ij} = - {v^2 \sin^2\beta }
 \lambda_{ik}^T M_N^{-1} \lambda_{kj}.
\label{tree}
\eeq
This tree level contribution provides a useful perturbation to the 1-loop
structure and provides the solar neutrino mass scale in this case. Combining
these two sources of neutrino mass we can
produce neutrino masses with a normal hierarchy. Assuming reasonable
values for the various g and h coefficients (which can be complex)
and with $a \sim b \sim 0.4$, $\ep \sim 0.20$
it is possible to achieve mass splittings compatible with measured
values (an appropriate diagonalisation procedure for a hierarchical
mass matrix is outlined in Ref.\cite{kingmns}). Due to the large value
of the (2,2) component of $m_\nu^{\rm
 loop}$ compared to the value of the (1,1) component, we do not
naturally produce large values for $\theta_{12}$. Consequently we need 
to moderately fine tune some of the g and h coefficients in order to
produce consistent mixing angles. Assuming the
mixing angles from the charged lepton sector are small, the
resulting MNS mixing angles produced from the neutrino sector can
accommodate the oscillation data. The analysis given in Ref.
\cite{kr} suggests small corrections from the charged lepton sector
are possible within the SU(3) flavour scenario.

\subsection{$\tev$ scale Leptogenesis from Susy breaking}
In this model we have large tri-linear scaler A-terms and consequently
the RH sneutrinos will be in deep thermal equilibrium at a scale
$\sim M_{\tilde{N_i}}$. Therefore the decay of the sneutrinos cannot
lead to the creation of a large asymmetry. The RH neutrinos on the
other hand are not in the thermal equilibrium due to the tiny
effective Yukawa couplings. In addition, the tree level vertex
diagram for the decay of the RH neutrinos is negligible compared to
the self energy diagram shown in Fig.1 of Ref.\cite{hmw}, which is
responsible for the asymmetry. Although the
diagram is suppressed by the Yukawa couplings it is enhanced by a
resonance effect when the mass splittings are naturally tiny as they
are for two of the RH neutrinos in the SO(3) model described in
this letter.
The form of the total asymmetry is \cite{Pil1,Pil2,HLNPS},
\begin{equation}
\varepsilon_{tot}=\sum_i\varepsilon_i=\sum_i\left(-\sum_{j \neq i}
  \frac{M_i }{M_j }\frac{\Gamma_j }{M_j }
  I_{ij} S_{ij}\right)  \,, \label{epsN}
\end{equation}
where
\begin{equation}
I_{ij} = \frac{ \hbox{Im}\,[ (\lambda^{(1)}  \lambda ^{(1)\dagger})_{ij}^2 ]}
{|\lambda^{(1)} \lambda ^{(1)\dagger} |_{ii} |\lambda^{(1)} \lambda ^{(1)\dagger} |_{jj}}
 \, ,\qquad
S_{ij} = \frac{M^2_j  \Delta M^2_{ij}}{(\Delta M^2_{ij})^2+M_i ^2
   \Gamma_j ^2} \, ,\qquad
\Gamma_j = \frac{|\lambda^{(1)} \lambda ^{(1)\dagger} |_{jj}}{8\pi}M_{j}
\,.
\label{selfenergy}
\end{equation}
Where  $\la^{(1)}=U_N \la$ are the 1-loop corrected Yukawa
couplings \footnote{Resummations of the Yukawa
  couplings have not been performed for simplicity, an example of such
  a procedure in the context of resonant leptogenesis is given in
  Ref.\cite{krystoff}.} with $U_N$ the unitary matrix that
diagonalises the full contribution to the RH neutrino mass matrix, \beq
M_N^R=M_N + \beta(M_N\la \la^\dagger  + \la^* \la^T M_N) +
\gamma\Delta M_{N} \label{mntot}\eeq
where\footnote{Note that the definitions of $\be$ and $\ga$ are
  modified compared to those given in Ref.\cite{mw}.}
\beq \beta \sim \frac{m_{3/2}}{h M_P} \biggl( \frac{g^2}{16 \pi^2}
\log{\frac{M_P}{M_N}} \biggr) \sim 10^{-15} \eeq and \beq
\gamma=\frac{m_{3/2}^2}{M_P} \sim 10^{-12}. \eeq Diagonalising
$M_N^R$ gives a mass splitting in the first two generations that is
the same parametric size as the width for these states. This
produces a resonance in the propagator of the virtual RH neutrino in
the self energy diagram for $N_1$ and $N_2$. This does not happen
when $N_3$ is present due to the much larger mass splitting between
$N_3$ and the other RH neutrino generations. Consequently, we only
get two pieces contributing significantly to $\varepsilon_{tot}$,
\beq \varepsilon_{tot}\simeq
  \frac{M_1 }{M_2 }\frac{\Gamma_2 }{M_2 }
  I_{12} S_{12}+  \frac{M_2 }{M_1 }\frac{\Gamma_1 }{M_1 }
  I_{21} S_{21}
\eeq
rearranging to give
\beq
\varepsilon_{tot}\simeq\frac{M_1M_2 I_{12}}{8 \pi}\Delta M^2_{12}\left[
    \frac{|\la^{(1)}\la^{\dagger(1)} |_{22} }{(\Delta M^2_{12})^2+M_1^2\Gamma_2
  ^2}+\frac{|\la^{(1)}\la^{\dagger(1)} |_{11} }{(\Delta M^2_{12})^2+M_2
  ^2\Gamma_1 ^2}\right].
\eeq
Using the same coefficients that were used to construct the
neutrino sector we find that we are actually a little bit off
resonance, such that $(\Delta M^2_{12})^2 > M^2\Gamma^2$. The actual
size of the mass splitting is of the order $\sim 10^{-8}\gev^2 $. This
is a little bigger than we might expect from the parametric sizes of the
non-diagonal RH neutrino contributions in eq. (\ref{mntot}). The large
size is due to the large mixing angle generated in the 1st two
generations as a result of the high degree of degeneracy in the masses
at tree level. We also have that
$|\la^{(1)}\la^{\dagger(1)} |_{22} \sim |\la^{(1)}\la^{\dagger(1)}
|_{11}$. Applying this we have,
\beq
\varepsilon_{tot}\simeq\frac{M_1M_2 I_{12}}{4 \pi}
    \frac{|\la^{(1)}\la^{\dagger(1)} |_{22} }{\Delta M^2_{12}}.
\eeq
Inserting, $\Delta M^2_{12} \sim 10^{-8} \gev ^2$,
$|\la^{(1)}\la^{\dagger(1)} |_{22} \sim 10^{-14}$ and $M_i \sim 10^2 \gev$
we have
\beq
\varepsilon_{tot}\sim I_{12}10^{-2}.
\eeq
The off-diagonal parts of $\la^{(1)}\la^{\dagger(1)}$, with these
parameters, are small compared to the diagonal parts due to
non-trivial cancellations, consequently, $I_{12}$ comes out to be of order
$10^{-5}$, giving,
\beq
\varepsilon_{tot}\sim 10^{-7}.
\eeq
Due to the sizes of the Yukawa couplings the decay
widths of the RH neutrinos are less than the Hubble constant and
therefore will not induce any wash-out effects via decays or
scatterings. The large A-terms do not contribute to any wash-out
effects as they need to be accompanied by a Yukawa interaction or a
lepton number violating B-term interaction (which is also small) in
order to break lepton number.
Thus with $g^* \sim 100$, $n_L/s$ can be of order
$\varepsilon_{tot}/100\sim 10^{-9}$ which is at the correct order to
give the CMBR-determined experimental value,
$n_B/n_\gamma=6.1^{+0.3}_{-0.2} \cdot 10^{-10}$ \cite{WMAP}.

\section{Conclusions}
In the context of supersymmetric theories, a weakly broken gauged SO(3)
flavour symmetry was used to produce two highly degenerate RH
neutrinos. It was shown that this SO(3) flavour
symmetry is compatible with all fermion masses and mixings if it is
supplemented with a further SU(3) flavour symmetry. A specific susy
breaking model was then used to generate the light neutrino masses as
well as a natural model of $\tev$ scale resonant leptogenesis. It must be
noted that this SO(3) flavour symmetry and its associated flavon field
alignments can be used independently of the susy breaking model used
to produce the neutrino masses in this letter. An application of this
was given in section 2 where degenerate RH neutrinos were generated in
the context of an effective superpotential.

\vskip 0.05in
\begin{center}
{\bf Acknowledgments}
\end{center}
\vskip0.05in
I wish to thank Thomas Hambye and especially John March-Russell and
Graham Ross for extremely useful discussions. This work is supported
by PPARC Studentship Award PPA/S/S/2002/03530.

\renewcommand{\theequation}{A-\arabic{equation}}
% redefine the command that creates the equation no.
\setcounter{equation}{0}  % reset counter
\section*{Appendix A}  % use *-form to suppress numbering
Below we list the assignments of all the fields in the theory.
\begin{table}[h]
\begin{center}
\begin{tabular}{|l|c|c|c|c|c|c|}
\hline Field & R-Charge & R-Parity & Z$_2$ & SU(3) & SO(3) & VEV
\\ \hline \hspace{1mm} $\ze^T$ & $-$1/5 & $+$ & $-$ & 1 & 3 & ($A$,$iA$,0) \\

\hspace{1mm} $\xi^T$ & $-$4/5 & $+$ & $+$ & 1 & 3 & ($D$,-$iD$,0) \\

\hspace{1mm} $\phi_3^T$ & 1 & $+$ & $+$ & \={3} & 1 & (0,0,$a_3$) \\

\hspace{1mm} $\phi_{23}^T$ & 1 & $+$ & $-$ & \={3} & 1 & (0,$b$,$b$) \\

\hspace{1mm} $\bar{\phi}_2$ & 0 & $+$ & $+$ & 3 & 1 & (0,$a_2$,0) \\

\hspace{1mm} $\bar{\phi}_3$ & $-$2 & $+$ & $+$ & $3$ & 1 & (0,0,$a_3$) \\

\hspace{1mm} $\bar{\phi}_{23}$ & 0 & $+$ & $+$ & $3$ & 1 & (0,$b$,$-b$) \\

\hspace{1mm} T & 4/3 & $+$ & $+$ & 1 & 1 &
(Fcpt,Acpt)=($m_I^2f_t$,$m_I$$a_t$) \\

\hspace{1mm} $Z_1$ & 23/15 & $+$ & $+$ & 1 & 1 &
(Fcpt,Acpt)=(0,$m_I$$a_{z1}$) \\

\hspace{1mm} $Z_2$ & 32/15 & $+$ & $-$ & 1 & 1 &
(Fcpt,Acpt)=(0,$m_I$$a_{z2}$) \\

\hspace{1mm} $T$ & 12/5  & $+$ & $+$ & 1 & 1 &
- \\

\hspace{1mm} $P$ & 18/5 & $+$ & $+$ & 1 & 1 & - \\

\hspace{1mm} $N$ & 2/3 & $-$ & $+$ & 1 & 3 & - \\

\hspace{1mm} $L$ & 4 & $-$ & $+$ & 3 & 1 & - \\

\hspace{1mm} $Q$ & 0 & $-$ & $+$ & 3 & 1 & - \\

\hspace{1mm} $U^c$ & 0 & $-$ & $+$ & 3 & 1 & - \\

\hspace{1mm} $D^c$ & 0 & $-$ & $+$ & 3 & 1 & - \\

\hspace{1mm} $E^c$ & $-$4 & $-$ & $+$ & 3 & 1 & - \\

\hspace{1mm} $H_u$ & 0 & $+$ & $+$ & 1 & 1 & v$_2$ \\

\hspace{1mm} $H_d$ & 0 & $+$ & $+$ & 1 & 1 & v$_1$ \\ \hline

\end{tabular}
\end{center}

\caption{Table of field assignments}
\label{tab1}
\end{table}

\renewcommand{\theequation}{B-\arabic{equation}}
% redefine the command that creates the equation no.
\setcounter{equation}{0}  % reset counter
\section*{Appendix B}  % use *-form to suppress numbering

Due to the R-charge assignments of the SO(3) flavon fields there are terms
that can be written down in addition to those in equation
(\ref{PT}). The additional terms are, \beq
PT\frac{(\ze_i\xi_i)^4}{M^7}+
PT\frac{(\ze_i \xi_i)^2 (\ze_j \ze_j)(\xi_k \xi_k)}{M^7}+ PT
\frac{(\ze_j \ze_j)^2(\xi_k \xi_k)^2}{M^7}+ T \frac{(\ze_j
\ze_j)^4}{M^6}. \eeq Along the F-flat direction
$|F_P|^2=0$, we now have, \beq  \vev{\ze_i\ze_i} +  \frac{\vev
{T(\ze_i\xi_i)^4}}{M^7} +  \frac{\vev {T(\ze_i\xi_i)^2(\ze_j
\ze_j)(\xi_k \xi_k)}}{M^7}+
 \frac{\vev{T(\ze_j\ze_j)^2(\xi_k \xi_k)^2}}{M^7}=0
\eeq
leading to $\vev {\ze^2} =0$ and $\vev T = 0$. Along the F-flat
direction $|F_T|^2=0$ applying $\vev {\ze^2} =0$ we have the condition,
\beq
 \vev{\xi_i\xi_i} +  \frac{\vev {P(\ze_i\xi_i)^4}}{M^7}=0
\eeq
leading to $\vev {\xi^2} =0$ and $\vev P = 0$, which are the
conditions we require for the correct vacuum alignment.

\end{document}